\documentclass[showpacs,amsmath,amssymb,nofootinbib,10pt]{revtex4}
\usepackage{graphicx}% Include figure files
\usepackage{dcolumn}% Align table columns on decimal point
\usepackage{bm}% bold math
\usepackage{color} %color

\begin{document}
\title{S and T Self-dualities in Dilatonic f(R) Theories}% Force line breaks with \\

\author{ Tongu{\c c} Rador}
\affiliation{
Department of Physics, Bo{\u g}azi{\c c}i University, 
Bebek TR34342, {\.I}stanbul, Turkey
}

\begin{abstract}
We search for theories, in general spacetime dimensions, that would incorporate a dilaton and higher powers of the scalar Ricci curvature such that they have exact S and/or T self-dualities. The theories we find are free of Ostrogradsky instabilities. We also 
show that within the framework we are confining ourselves a theory of the form mentioned
above can not have both T and S dualities except for the case where the action is linear in the scalar curvature.
\end{abstract}

\maketitle

\section{Introduction}

Theories that contain higher powers of the curvature scalar has atracted much attention over the past years, after their introduction some time ago \cite{o1}-\cite{o2}, espescially in view of the accelerated expansion of the universe and the possible avenues related to extra dimensions inspired by string theory. The litterature became  too voliminous to cite even partially, so we refer to the reviews on the subject \cite{rev1}-\cite{rev5} and the references therein.

As well known the low energy string theory action is that of Einstein gravity with a non-minimally coupled dilaton. This lowest order action has two important symmetries: T and S dualities. In this work we investigate the conditions on how to have S and/or T symmetries in  and $f(R)$ theory, as higher order corrections to the string action will surely involve higher orders of the curvature objects. The reason to confine the study to actions that contain only powers of the Ricci curvature scalar is the fact that pure $f(R)$ theries are free from Ostrogradsky instabilities \cite{os3}; as a consequence related to this the models we find are
also free of the mentioned problem, albeit in a slightly disguised manner.

\section{An Observation}

Let us consider the following action in $d$ dimensions

\begin{equation}{\label{action1}}
\int dx^{d} \sqrt{-\tilde{G}}\; e^{-(\frac{d-2}{4})\tilde{\phi}}\; \left[ \tilde{R}+A(\tilde{\nabla}\tilde{\phi})^{2}\right]\;,
\end{equation}
\noindent where $A$ is a real number. It can be shown that the action is form invariant under the following transformations
\begin{subequations}\label{sdualt}
\begin{eqnarray}
\tilde{G}_{\mu\nu} &=& e^{-\phi}G_{\mu\nu}\;,\\
\tilde{\phi}&=&-\phi\;,
\end{eqnarray}
\end{subequations}
\noindent which are called S duality transformations. If the action is form invariant one may call the model S self-dual.

However, to achieve this conclusion one has to perform an integration by parts and hence convert some of the terms to boundary integrals which in turn, as well known, will not effect the Euler-Lagrange equations of motion since to find them one assumes  that the variations vanish at the boundary.

Furthermore as is also well known, one can represent the theory in the Einstein frame as opposed to the original Jordan frame by the following conformal transformation of the metric

\begin{equation}\label{eq:jortoe}
\tilde{G}_{\mu\nu} = e^{\tilde{\phi}/2} G_{\mu\nu}.
\end{equation}

This will transform the original action into the following

\begin{equation}
\int dx^{d}\sqrt{-G}\left[ R+(A-\frac{(d-1)(d-2)}{16})\left(\nabla\tilde{\phi}\right)^{2}-\frac{(d-1)}{2}\nabla^{2}\tilde\phi\right].
\end{equation}
where the non-tilde derivatives are those related to $G$. Here we also realize the last term as a surface term and it can thus be ignored, if one would only like to get equations of motion. Furthermore if one also requires
the scalar field kinetic term to be normalized to the canonical value of $-1/2$ one requires $A=4$ for $d=10$; the canonical numbers of low energy string theory. Surely the remnant of the original symmetry presents itself here as the $\tilde{\phi}\to-\tilde{\phi}$ invariance.

We see that in both instances a manipulation of surface terms is necessary. However this will not  necessarily be possible if one has a theory that involves higher powers of $R$ and this may impede S self-duality of the theory.

\section{S self-dual theories}

First of all let us note that one has

\begin{subequations}{\label{alg1}}
\begin{eqnarray}
\left[\tilde{R}+\frac{(d-1)}{2}\tilde{\nabla}^{2}\tilde{\phi}\right]&=& e^{\phi}\left[R+\frac{(d-1)}{2}\nabla^{2}\phi\right]\\
A(\tilde{\nabla}\tilde{\phi})^{2}&=& e^{\phi} A (\nabla\phi)^{2}
\end{eqnarray}
\end{subequations}
under the S duality transformations. These algebraic properties can be exploited to form theories that will involve higher powers of $R$ in such a way that the theory can be made S self-dual.

In fact one can further generalize the possible terms in (\ref{alg1}) by multiplying them with functions which are even in $\tilde{\phi}$. That is, under (\ref{sdualt}), one
still has
\begin{subequations}{\label{alg2}}
\begin{eqnarray}
\alpha(\tilde{\phi})\left[\tilde{R}+\frac{(d-1)}{2}\tilde{\nabla}^{2}\tilde{\phi}\right]&=& e^{\phi}
\alpha(\phi)\left[R+\frac{(d-1)}{2}\nabla^{2}\phi\right]\\
A(\tilde{\phi})(\tilde{\nabla}\tilde{\phi})^{2}&=& e^{\phi} A(\phi) (\nabla\phi)^{2}
\end{eqnarray}
\end{subequations}
provided that $\alpha(-\phi)=\alpha(\phi)$ and $A(-\phi)=A(\phi)$. Furthermore one can also add a pure potential term for the scalar field which has the same algebraic transformation rule

\begin{equation}
e^{-\tilde{\phi}/2}V(\tilde{\phi})=e^{\phi}\;e^{-\phi/2}V(\phi)
\end{equation}
provided again that one has $V(-\phi)=V(\phi)$.

Thus the most general term that has the same algebraic transformation rule is the following

\begin{equation} 
\mathcal{U}_{i}(\tilde{G}_{\mu\nu},\tilde{\phi})\equiv \alpha_{i}(\tilde{\phi})\left[\tilde{R}+\frac{(d-1)}{2}\tilde{\nabla}^{2}\tilde{\phi}\right]+A_{i}(\tilde{\phi})(\tilde{\nabla}\tilde{\phi})^{2}+e^{-\tilde{\phi}/2}V_{i}(\tilde{\phi})
\end{equation}
in that it becomes
\begin{equation}
\mathcal{U}_{i}(\tilde{G}_{\mu\nu},\tilde{\phi})=e^{\phi}\mathcal{U}_{i}(G_{\mu\nu},\phi)
\end{equation}
under the S duality transformations (\ref{sdualt}), provided $\alpha_{i}$, $A_{i}$ and $V_{i}$ are all even functions.

Now let us  define the following object
\begin{equation}
M(\tilde{G},\tilde{\phi})\equiv \sqrt{-\tilde{G}}e^{-(d-2)\tilde{\phi}/4}
\end{equation}
which transforms as
\begin{equation}
M(\tilde{G},\tilde{\phi})=e^{-\phi}M(G,\phi)
\end{equation}

One can show that the following object

\begin{equation}
\mathcal{L}^{(n)}(\tilde{G}_{\mu\nu},\tilde{\phi})\equiv M(\tilde{G},\tilde{\phi})e^{(n-1)\tilde{\phi}/2}\prod_{a=1}^{n}\mathcal{U}_{i_{a}}(\tilde{G}_{\mu\nu},\tilde{\phi})
\end{equation}
is S self-dual in the algebraic sense mentioned above. That is one has

\begin{equation}
\mathcal{L}^{(n)}(\tilde{G}_{\mu\nu},\tilde{\phi})=\mathcal{L}^{(n)}(G_{\mu\nu},\phi)
\end{equation}
under (\ref{sdualt}).

So we have arrived at the most general S self-dual action that contains higher powers of the scalar curvature

\begin{equation}{\label{laggen}}
S=\int dx^{d} \sum_{n}\mathcal{L}^{(n)}(\tilde{G}_{\mu\nu},\tilde{\phi}).
\end{equation}

\subsection{Ostrogradski Instabilities}

Whenever one has a theory with a lagrangian that contains higher derivative terms one is facing the troublesome Ostrogradski instabilities. Since the S self-dual theories we have introduced has a generic lagrangian which depends on the second derivatives of the dilaton field  we have to assess if one truly has  this problem.

Now it is a well known fact that pure $f(R)$ theories are free of Ostrogradski instabilities even though they contain higher derivatives of the metric. On top of this fact let us switch to the Einstein frame via (\ref{eq:jortoe}). This will result in the following

\begin{equation}
\mathcal{U}_{i}(\tilde{G}_{\mu\nu},\tilde{\phi})=e^{-\tilde{\phi}/2}\mathcal{V}_{i}(G_{\mu\nu},\tilde{\phi}),
\end{equation}
where

\begin{equation}\label{eq:gensdual}
\mathcal{V}_{i}(G_{\mu\nu},\tilde{\phi})=\alpha_{i}(\tilde{\phi})R+\left[A_{i}(\tilde{\phi})+\alpha_{i}(\tilde{\phi})\frac{(d-1)(d-2)}{8}\right]\left(\nabla\tilde{\phi}\right)^{2}+V_{i}(\tilde{\phi}).
\end{equation}

Consequently original lagrangians of the form (\ref{laggen}) do not have higher derivatives of the scalar field in the Einstein frame. We therefore conclude that the theories we have introduced will be free of Ostrogradskian instabilities. 

One can understand this absence also in the original Jordan frame. There, one has  double derivatives of the scalar field in the action but they always come accompanied by the curvature term in the form $\tilde{R}+(d-1)\tilde{\nabla}^{2}\tilde{\phi}/2$
and thus always get mixed with the degrees of the geometry and hence do not create a truly independent canonical momentum.

The fact that Ostrogradskian instabilities are absent for the dilaton field does not necessarily mean that none of the equations of motion involves higher derivatives of it. It simply means that there is at least an equation of motion that depends only on its derivatives of second or first degree. To expose this let us evaluate the full equations of
motion in the Einstein frame. The action becomes

\begin{equation}
\int dx^{d}\sqrt{-g}f(\mathcal{V})
\end{equation}
with
\begin{equation}
\mathcal{V}=\alpha(\chi)R+\beta(\chi)\left(\nabla\chi\right)^{2}+V(\chi)
\end{equation}
with $\beta(\chi)=\left[A(\chi)+\alpha(\chi)(d-1)(d-2)/8\right]$. We have also set $\tilde{\phi}=\chi$ to get rid of the tildes.

One can show that the variation of the action with respect to the metric gives the following equation,

\begin{equation}{\label{einseqR}}
\alpha f' R_{\mu\nu}-\frac{1}{2}g_{\mu\nu}f+(g_{\mu\nu}\nabla^{2}-\nabla_{\mu}\nabla_{\nu})(\alpha f')+\beta f' (\nabla_{\mu}\chi)(\nabla_{\nu}\chi)=0
\end{equation}
where primes indicate derivatives with respect to the full  argument of the corresponding function. This equation obviously incorporates derivative of the scalar field higher than second order. But the variation of the action with respect to the scalar field brings

\begin{equation}{\label{einseqS}}
\left[\alpha' R+\beta' (\nabla\chi)^{2}+V'\right]f'=\nabla_{\mu}\left[2\beta f'\nabla^{\mu}\chi\right]
\end{equation}

It is manifestly clear from the above that the equations of motion for the scalar field involves only up to and including the second derivative and thus any higher order derivative is fixed by
this equation.
\subsection{Pure f(R) theories in the Einstein Frame}

One by product of our discussion is that a pure $f(R)$ theory can be thought of as an Einstein frame representation of an S self-dual dilatonic theory provided one has all $\alpha_{i}$, $A_{i}$ and $V_{i}$ as constants and satisfying  the following condition
\begin{equation}\label{asker0}
A_{i}+\alpha_{i}\frac{(d-1)(d-2)}{8}=0
\end{equation} 

This condition is quite non-trivial in the sense that it binds the coefficients of the
scalar curvature to those of the coefficient of the kinetic term for the scalar field in the Jordan frame. The resulting theory in the Jordan frame, though by construction S self-dual, may have tachyonic instabilities due a wrong signature. 

In fact one can envisage a slightly more general framework. If we require an originally
self S-dual theory to end up a pure f(R) theory in the Einstein frame one does not
necessarily need to require the coefficients $\alpha_{i}$, $A_{i}$ and $V_{i}$ to be constants. We may still require them to be functions of the field $\tilde{\phi}$. In this
case the condition above becomes

\begin{equation}{\label{asker1}}
A_{i}(\tilde{\phi})+\alpha_{i}(\tilde{\phi})\frac{(d-1)(d-2)}{8}=0
\end{equation}

These  conditions of the field $\tilde{\phi}$ should be simultaneously satisfied. The simplest way for this to happen is to assume that the theory in the original Jordan frame takes, modulo the measure term, the form of $f(\mathcal{U})$ where there is only one $A$, $\alpha$ and $V$ function;

\begin{equation}
\mathcal{U}=\alpha(\tilde{\phi}) (\tilde{R}+\frac{d-1}{2}\tilde{\nabla}^{2}\tilde{\phi})+A(\tilde{\phi})(\tilde{\nabla}\tilde{\phi})^{2}+e^{-\tilde{\phi}/2}
V(\tilde{\phi})
\end{equation}
and (\ref{asker1}) becomes a single  equation and may more easily be accomodated. Thus in this generalized case the dilaton in the Jordan frame must have already been stabilized in a subtle way: A quite different condition than (\ref{asker0}).

\section{T duality}

Let us go back to the original action (\ref{sdualt}) and study the effect of T-duality on it. T-duality is a symmetry that acts on the internal dimensions; so a separation of what
is internal or not is mandatory. The usual and simplest way to incorporate it is to assume the following separation in metric;

\begin{equation}{\label{metan}}
ds^{2}=\tilde{G}_{\mu\nu}dX^{\mu}dX^{\nu}=g(x)_{mn}dx^{m}dx^{n}+e^{2\tilde{C}(x)}\gamma(y)_{ij}
dy^{i}dy^{j}
\end{equation}
which can be called the {\em compactification} or {\em warped} ansatz: The metric is assumed to be block diagonal with respect to the co-ordinates $y$ of the compactified extra dimensions and the co-ordinates $x$ of the observed dimensions, which include time. Here $\tilde{C}(x)$ is called the radion field and $\gamma(y)_{ij}$ is a metric for the manifold of extra dimensions which does not depend on the co-ordinates of the observed dimensions which also include the time variable. We set the dimensionality of the extra dimensions to be $p$. In view of this ansatz we also assume that the dilaton field $\tilde{\phi}$ does not depend on the co-ordinates of the internal dimensions.

This ansatz allow us to evaluate the Ricci scalar of the full spacetime as follows \cite{yam1}-\cite{yam2},
\begin{equation}{\label{TdRR}}
\tilde{R}=R[\tilde{G}]=R[g]+e^{-2\tilde{C}}\tilde{R}[\gamma]-2 p e^{-\tilde{C}}g^{mn}\nabla_{m}\nabla_{n}e^{\tilde{C}}-p(p-1)e^{-2\tilde{C}}g^{mn}(\nabla_{m}e^{\tilde{C}})(\nabla_{n}e^{\tilde{C}})
\end{equation}
where the covariant derivatives $\nabla$ are now those of the metric $g_{mn}$. We also have 
\begin{equation}{\label{Tdmes}}
\sqrt{-\tilde{G}}\;e^{-(d-2)\tilde{\phi}/4}=\sqrt{-g}\sqrt{\vert\gamma\vert}\;e^{p\tilde{C}-(d-2)\tilde{\phi}/4}.
\end{equation}
and
\begin{equation}
A(\tilde{\nabla}\tilde{\phi})^{2}=A\tilde{G}^{\mu\nu}(\tilde{\nabla}_{\mu}\tilde{\phi})
(\tilde{\nabla}_{\nu}\tilde{\phi})=Ag^{mn}(\nabla_{m}\tilde{\phi})(\nabla_{n}\tilde{\phi})
\end{equation}
following directly from the ansatz.

Now let us define the following transformation on these variables,

\begin{subequations}
\begin{eqnarray}
\tilde{C}&=&-C\\
\tilde{\phi}&=&\phi-\frac{8p}{d-2}C
\end{eqnarray}
\end{subequations}
which automatically leave the measure  in (\ref{Tdmes}) invariant. These are called T-duality transformations and one can show that the action (\ref{action1}) is form invariant under them provided one fixes

\begin{subequations}{\label{TSD0}}
\begin{eqnarray}
A &=& (\frac{d-2}{4})^{2}\longrightarrow A=4\;{\rm for}\;d=10\\
R[\gamma]&=&0
\end{eqnarray}
\end{subequations}

However one must realize, in view of the form in (\ref{TdRR}) which involves double derivatives of $\tilde{C}$, that again a manipulation involving integration by parts is required. So in conclusion the action in (\ref{action1}) is both S and T self-dual provided the coefficient of $A$ is fixed to be predetermined constant in (\ref{TSD0}). These are of course well known properties of the low energy action of string theory. The general point we make is that manipulations of it via integration by parts is necessary for both symmetries. We were able to generalize S self-duality to actions that are required to involve higher powers of the scalar curvature. We now look for generalizing T self-duality.

\subsection{T self-dual theories}

Let us remember that the measure prefactor is invariant under the T-duality transformations. Thus, it is clear that  we need to add a term to $R[\tilde{G}]$ that would compensate the appearance of the double derivative of $\tilde{C}$, if we are to
get an algebraic invariance similar to what happened in the generalization of S-duality transformations.
 
The only term that can be added in the original variables is the following
\[
\tilde{G}^{\mu\nu}\tilde{\nabla}_{\mu}\tilde{\nabla}_{\nu}\tilde{\phi}.
\] 
Using the compactification
ansatz (\ref{metan}) we can simply infer that the inverse metric $G^{\mu\nu}$ is also
block diagonal. This along the assumption that the dilaton field is blind to co-ordinates
$y^{i}$ of the extra dimensional metric means that one has
\begin{equation}
\tilde{G}^{\mu\nu}\tilde{\nabla}_{\mu}\tilde{\nabla}_{\nu}\tilde{\phi}=g^{mn}\left(
\nabla_{m}\nabla_{n}\tilde{\phi}+p\nabla_{m}{\tilde{C}}\nabla_{n}{\tilde{\phi}}\right)
\end{equation}
where the covariant derivatives on the right hand side are those related to the metric $g_{mn}$. With these considerations, after a slightly tedious calculation, one can show that the
following object
\begin{equation}
R[\tilde{G}]+A(\tilde{\nabla}\tilde{\phi})^{2}+\sigma \tilde{\nabla}^{2}\tilde{\phi}
\end{equation}
after the metric compactification ansatz is T self-dual provided the following conditions
are met
\begin{subequations}{\label{TSdcond}}
\begin{eqnarray}
A&=&-\left(\frac{d-2}{4}\right)^{2}\\
R[\gamma]&=& 0\\
\sigma &=&\frac{d-2}{2}
\end{eqnarray}
\end{subequations}

The difference in the value of $A$ in the above as opposed to the one in (\ref{TSD0}) is
to be attributed to the fact that in (\ref{TSD0}) an integration by parts is incorporated
to show the T self duality of the action linear in $R$. Thus they are not actually different conditions if one considers only the action (\ref{action1}). We also see that in view of its somewhat contrived form, T-duality forbids a potential term for the dilaton.

\subsection{Incompatibility of T and S self-dualities for $f(R,\phi)$ theories}

Note that the coefficient of the term $\tilde{\nabla}^{2}\tilde{\phi}$ are found to be
different for S self-duality and T self-duality conditions: $(d-1)/2$ in the former case
and $(d-2)/2$ in the latter. Fixing  the functions $\alpha_{i}(\tilde{\phi})$ and $A_{i}(\tilde{\phi})$ to be constants and $V_{i}(\tilde{\phi})=0$ in (\ref{eq:gensdual}) is necessary of course but can not save the situation. The reason for this discrepancy can be attributed to the fact that S duality transformations act on the full space-time defined by the general metric $\tilde{G}_{\mu\nu}$ whereas for T duality transformations one must first assume
the compactification ansatz. These two are very different actions and the dilaton simply becomes overworked especially in the absence of help from surface integral simplification strategies. Using two dilaton fields, one responsible for T-duality one for S-duality does not work either. Nor contemplating a workaround by the 3-form field of string theory will help since it can not produce a Laplacian of the dilaton, to compensate the mismatch. 

Furthermore this is not the whole issue. Even we could have found an algebraic TS self
dual combination of fields which is linear in the curvature the form of S duality we found for theories that involve higher powers of the curvature needs powers of the exponential of the dilaton to compensate the conformal factors arising from the transformations. But a linear term is not our aim. Thus;

{\em There are no both S and T self-dual categories of theories, in the algebraic sense defined in this work, which are free of Ostrogradski instabilities containing a single dilaton field and incorporating higher powers of the scalar curvature.}

However, we must remind the reader, that for instance relaxing the condition on the non-existance of Ostrogradski instabilities
would yield theories that involve higher powers of curvature objects, not necessarily the Ricci scalar, and still incorporate
both T and S dualities.


\begin{thebibliography}{100}
\bibitem{o1} H.~A.~Buchdahl, Monthly Notices of the Royal Astronomical Society. 150: 1–8, 1970.
\bibitem{o2} A.~A.~Starobinsky, Phys. Lett. B. 91: 99–102, 1980.
\bibitem{rev1} A.~De Felice and S.~Tsujikawa, Living Rev. Relat. 13, 3, 2010.
\bibitem{rev2} T.~P.~Sotiriou and V.~Faraoni, Rev. Mod. Phys. 82, 451-497, 2010.
\bibitem{rev3} S.~Capozziello and M.~De Laurentis, Phys. Rep. 509, 4-5: 167-321, 2011. 
\bibitem{rev4} S.~Capozziello and V.~Faraoni, {\em Beyond Einstein Gravity: A Survey of Gravitational Theories for Cosmology and Astrophysics}, Springer, 2011, ISBN: 978-94-007-0165-6.
\bibitem{rev5} V.~Faraoni, Volume 38 of the series Astrophysics and Space Science Proceedings, 19-32, 2013.
\bibitem{os1} M.~Ostrogradsky, Mem. Ac. St. Petersbourg VI 4, 385, 1850.
\bibitem{os3} R.~P.~Woodard, Lect. Notes. Phys. 720, 403, 2007.
\bibitem{os2} R.~P.~Woodard, arxiv:1506.02210v2, 2015.
\bibitem{yam1} F.~Dobarro and E.~Lami~Dozo, Trans. Amer. Math. Soc. 303, 161-168, 1987.
\bibitem{yam2} H.~Yamabe, Osaka Math. J. 12, 21-37, 1960.
\end{thebibliography}
\end{document}